\documentclass[letterpaper,aps,prd,preprint,showpacs,nofootinbib,superscriptaddress]{revtex4}

\usepackage{graphicx}  
\usepackage{amsfonts}

\begin{document}

\title{Prospects for Sneutrino Dark Matter in the BLSSM}
\author{Luigi Delle Rose}
\email[]{L.Delle-Rose@soton.ac.uk}
\affiliation{School of Physics and Astronomy, University of Southampton, Highfield, Southampton SO17 1BJ, United Kingdom}\affiliation{Particle Physics Department, Rutherford Appleton Laboratory, Chilton, Didcot, Oxon OX11 0QX, United Kingdom}

\author{Shaaban Khalil}
\email[]{Skhalil@zewailcity.edu.eg}
\affiliation{Center for Fundamental Physics, Zewail City of Science and Technology, Sheikh Zayed,12588 Giza, Egypt}

\author{Simon J.D. King}
\email[]{SJD.King@soton.ac.uk}
\affiliation{School of Physics and Astronomy, University of Southampton, Highfield, Southampton SO17 1BJ, United Kingdom}

\author{Suchita Kulkarni}
\email[]{suchita.kulkarni@oeaw.ac.at}
\affiliation{Institut f{\"u}r Hochenergiephysik, {\"O}sterreichische Akademie der Wissenschaften, Nikolsdorfer Gasse 18, 1050 Wien, Austria}

\author{Carlo Marzo}
\email[]{Carlo.Marzo@kbfi.ee}
\affiliation{National Institute of Chemical Physics and Biophysics, R{\"a}vala 10, 10143 Tallinn, Estonia}

\author{Stefano Moretti}
\email[]{S.Moretti@soton.ac.uk}
\affiliation{School of Physics and Astronomy, University of Southampton, Highfield, Southampton SO17 1BJ, United Kingdom}\affiliation{Particle Physics Department, Rutherford Appleton Laboratory, Chilton, Didcot, Oxon OX11 0QX, United Kingdom}

\author{Cem S. Un}
\email[]{cemsalihun@uludag.edu.tr}
\affiliation{Department of Physics, Uluda\~{g} University, TR16059 Bursa, Turkey}


\begin{abstract}
The ($B-L$) Supersymmetric Standard Model (BLSSM) motivates several Dark Matter (DM) candidates beyond the Minimally Supersymmetric Standard Model (MSSM). We assess the comparative naturalness of the two models and discuss the potential detection properties of a particular candidate, the Right-Handed (RH) sneutrino.
\end{abstract}
\maketitle

\section{Introduction}
In the SM there is a global ($B-L$) symmetry which is conserved. By extending the MSSM by gauging this symmetry, based on the group $SU(3)_c \times SU(2)_L \times U(1)_Y \times U(1)_{B-L}$, one finds a significantly enriched particle content. Firstly, when gauging $(B-L)$, one must add three SM singlet fields to cancel the triangle anomaly diagrams, which may be identified as the RH neutrinos. We combine this with a low (TeV) scale Type-I see-saw mechanism to explain the light, non-vanishing Left-Handed (LH) neutrino masses; something the MSSM does not do. In addition to these RH neutrinos, one may spontaneously break the ($B-L$) symmetry with new Higgses, called bileptons which carry a ($B-L$) charge of $\pm 2$. This will give rise to a new, massive, gauge boson associated with this group, a $Z'$. Finally, one finds the superpartners of these new particles, the RH sneutrinos, bileptinos and partner of the new $B'$ boson, the BLino.

In this work, we consider the implications of this extension to the MSSM. One finds all the benefits of the MSSM (gauge coupling unification, hierarchy problem solution, etc.) in addition to several new DM candidates and an explanation of light, non-vanishing neutrino masses. We compare the naturalness of the two models, the MSSM and BLSSM, in the presence of universal scalar and gaugino masses (often called the Constrained MSSM (CMSSM) scenario).

We proceed as follows. In section \ref{sec:Naturalness} we introduce the concept of naturalness and compare the non-minimal scenario to the MSSM. Then, we consider the various DM candidates of the model, and compare the relic abundances of the candidates for each model in section \ref{sec:DM}. For the remainder of this work, we discuss the ability to detect sneutrino DM via direct/indirect and collider detections, respectively, in sections \ref{sec:Indirect_Searches} and \ref{sec:Collider_Searches}. Finally, we conclude.

\section{Naturalness}
\label{sec:Naturalness}
Whilst there is no fundamental measure of naturalness in nature, to compare two models quantitatively one must use some metric. The weak scale ($M_Z$) depends on the soft Supersymmetry (SUSY) breaking terms through the Renormalisation Group Equations (RGEs) and the  Electro-Weak (EW) minimisation conditions, which can be expressed as 
\begin{equation}
\frac{1}{2} M_Z^2= \frac{m_{H_d}^2 - m_{H_u}^2 \tan^2 \beta}{\tan^2 \beta -1} -  \mu^2  . 
\label{EWmin}
\end{equation}
Because of this, we choose here to adopt the criterion of Fine Tuning (FT) related to a change in the $Z$-boson mass. We denote this as $\Delta$, defined by the largest change when altering parameters in the theory \cite{Ellis:1986yg,Barbieri:1987fn}
\begin{equation}
\Delta={\rm Max} \left| \frac{\partial \ln v^2}{\partial \ln a_i}\right| =  {\rm Max} \left| \frac{a_i}{v^2} \frac{\partial v^2}{\partial a_i } \right|  = {\rm Max} \left| \frac{a_i}{M_Z ^2} \frac{\partial M_Z ^2}{\partial a_i} \right|.
\label{eq:BGFT}
\end{equation}
A point in parameter space has a low FT if the SM $Z$ mass does not largely deviate when deviating from its position. A natural model will consequently posses large regions of viable parameter space with such points. The absolute scale of FT is largely irrelevant as $\Delta$ has no physical meaning, we thus are only concerned with the relative value between two models. In this work we consider the fundamental parameters of the theory to be the Grand Unification Theory (GUT) ones, though other works consider the FT at EW level as well \cite{Ross:2017kjc,DelleRose:2017ukx,DelleRose:2017smp,DelleRose:2017hvy}. The input GUT parameters are: the unification masses for scalars ($m_0$)  and gauginos ($m_{1/2}$), the universal trilinear coupling ($A_0$), the $\mu$ parameter and the quadratic soft SUSY term ($B\mu$),
\begin{equation}
a_i = \left\lbrace m_0 , ~ m_{1/2},~ A_0,~ \mu ,~ B \mu  \right\rbrace.
\end{equation}
For the BLSSM, one also has two further parameters, $(\mu',~ B \mu')$. We find that the dominating term for both the MSSM and BLSSM comes from $\mu$, which is fixed by EW Symmetry Breaking (EWSB). It is important to check the level of FT in the BLSSM, as one may have expected that a heavy $Z'$ would lead to a large $\mu '$ which could, in principle, have surpassed all other contributions, however we do not see such behaviour. We will now compare the FT for the two models.

The scan performed to obtain these data has been done by SPheno \cite{Porod:2003um} with all points being passed through \texttt{HiggsBounds} \cite{Bechtle:2008jh,Bechtle:2011sb,Bechtle:2013wla,Bechtle:2015pma} and \texttt{HiggsSignals} \cite{Bechtle:2013xfa}. We have scanned over the range $[0,5]$ TeV in both $m_0$ and $m_{1/2}$, $\tan \beta$ in $[0 , 60]$, $A_0$ in $[-15, 15]$ TeV, which are common universal parameters for both the MSSM and the BLSSM, while for the BLSSM we also required $\tan \beta'$ in the interval $[0,2]$ with neutrino Yukawa couplings $Y^{(1,1)}$, $Y^{(2,2)}$, $Y^{(3,3)}$ in $[0,1]$. The $M_{Z'}$ value has been fixed to 4 TeV to comply with collider searches \cite{Khachatryan:2016zqb}.

\begin{figure}[h]
\centering
\includegraphics[width=0.49\linewidth]{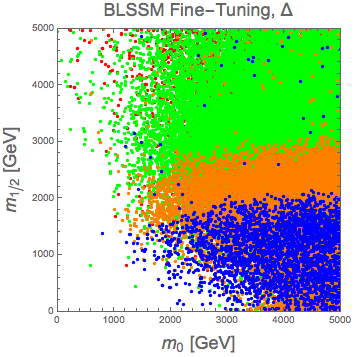}
\includegraphics[width=0.49\linewidth]{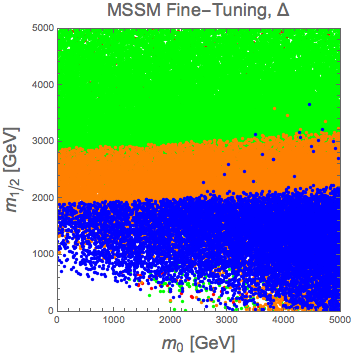}
\caption{Fine-tuning in the plane of unification of scalar, gaugino masses for BLSSM and MSSM for GUT-parameters ($\Delta$). The FT is indicated by the colour of the dots: blue for FT $<$ 500; Orange for 500 $<$ FT $<$ 1000; Green for 1000 $<$ FT $<$ 5000; and Red for FT $>$ 5000.}
\label{fig:BLSSMGUT-loop-m0m12}
\end{figure}

In figure \ref{fig:BLSSMGUT-loop-m0m12}, we compare the FT for the MSSM and BLSSM in the plane of ($m_0$,~$m_{1/2}$). We colour the parameter points according to their FT, $\Delta$: red for $\Delta > 5000$, orange for 500 $<$ $\Delta$ $<$ 1000 and blue (the least finely-tuned points) for $\Delta$ $<$ 500. Overall the picture between the two models is very similar. Firstly, we see there are few viable BLSSM parameter points for low $m_0$. Both models follow the general trend that the FT is strongly related to universal gaugino mass $m_{1/2}$, due to the $\mu$ dependence on this. Finally, we summarise that the MSSM and the BLSSM have very similar levels of FT.

\section{Relic Abundance}
\label{sec:DM}

\begin{figure}[h]
	\centering
	\includegraphics[width=0.49\linewidth]{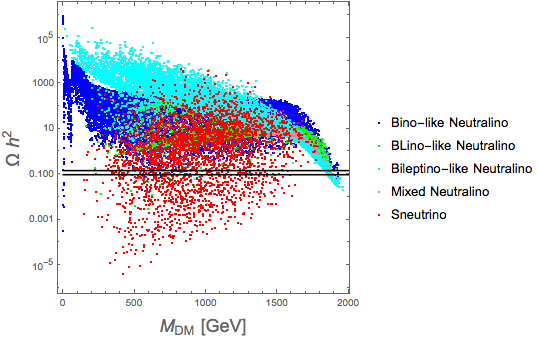}
	\includegraphics[width=0.49\linewidth]{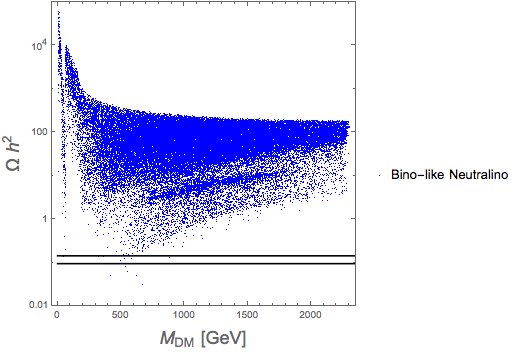}
	\caption{Left: relic density vs LSP mass for the BLSSM.   
		Right: relic density vs LSP mass for the MSSM. In both plots the horizontal lines identify the $2\sigma$ region around the current central value of  $\Omega h^2$. }
	\label{fig:BLSSM-DM}
\end{figure}

By our previous measure, $\Delta$, both the universal MSSM and BLSSM scenarios had a similar level of FT. Another measure of FT could be to compare the relative size of parameter space in both models, over the same ranges in GUT parameters. Before enforcing a DM candidate, in both cases the number of viable points (satisfying LHC searches and a SM-like Higgs) is a similar fraction of the total number with a uniform random scan. However, once we enforce that the LSP be the DM candidate and comply with the correct relic density, one finds that the BLSSM has significantly larger viable regions of parameter space.

In SUSY the Lightest Super Partner (LSP) is stable due to $R$-parity conservation. In the MSSM, this is imposed ad-hoc, but for the BLSSM this is automatically satisfied as the $R$-symmetry is defined as $R=(-1)^{3(B-L)+2S}$, and $B-L$ is conserved (or broken by two units, which leaves a residual $\mathbb{Z}_2$). In addition to the bino-like neutralino in the MSSM, there are three extra DM candidates in the BLSSM. Firstly, there are bileptino-like and BLino-like neutralinos, and also the superpartner of the RH neutrino, the RH sneutrino. 

Figure \ref{fig:BLSSM-DM} plots the relic density (with 2$\sigma$ bounds) \cite{Ade:2015xua} against the mass of the LSP ($M_\mathrm{DM}$). For the bino-like neutralino in the MSSM (the only allowed candidate due to other constraints in the universal scenario), very few of the parameter points survive. The vast majority here have a very large relic abundance. For the BLSSM, there are several bino-like neutralino LSP parameter points which satisfy the relic density, and furthermore one is guaranteed regions of parameter space where this is satisfied when $M_\mathrm{DM} \sim \frac{1}{2}M_{Z'}$, in our case around the 2 TeV mass. This effect also occurs for BLino and bileptino-like neutralinos. However, the RH sneutrinos offer the largest number of parameter points of our scan which satisfy the relic density requirements and we discuss the potential detection of this DM candidate in subsequent sections.

\section{Direct and Indirect Searches}
\label{sec:Indirect_Searches}

The mass eigenstate of the RH sneutrino LSP is CP-even (Re) or CP-odd (Im), due to lepton number violating operators. Consequently, two DM particles interacting must produce a CP-even state, in addition to conservation of angular momentum and spin statistics. One finds the largest cross section of annihilation into a pair of the lightest ($B-L$)-like CP-even Higgses, either through the four point $\left( \tilde{\nu}^{({\rm R,I})}_{{1}} \tilde{\nu}^{({\rm R,I})}_{{1}} \rightarrow h_i h_j \right)$ or mediated by a single CP-even Higgs $\left( \tilde{\nu}^{({\rm R,I})}_{{1}} \tilde{\nu}^{({\rm R,I})}_{{1}} \rightarrow  h_i \rightarrow h_i h_j \right)$ interaction. If this is not kinemetically available (ie $M_{h_2} > M_{\tilde{\nu}}$), then the next largest cross section is a decay to a charged $W^+ W^-$ pair. It is this decay which will provide a signal for indirect detection, which we discuss shortly. One also finds RH sneutrino interactions under the gauge boson associated with the $U(1)_{B-L}$ group, the $Z'$, but this requires one CP-even and one CP-odd sneutrino $\left( \tilde{\nu}^{({\rm R,I})}_{{1}} \tilde{\nu}^{({\rm I,R})}_{{1}} \rightarrow Z' \right)$. All of these processes require a heavy mediator and so are suppressed when it comes to Direct Detection (DD).

\begin{figure}[h]
	\centering
	\includegraphics[width=0.8\linewidth]{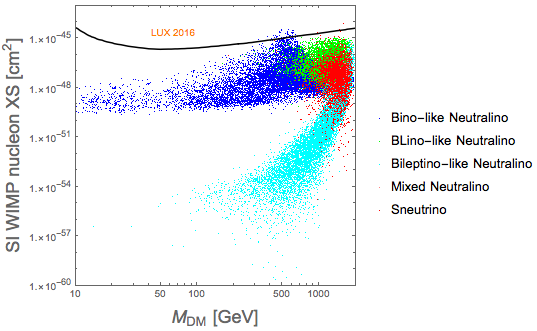}
	\caption{Spin-independent WIMP-nucleus scattering cross section generated in our scan against the upper bounds from 2016 run of the LUX experiment.}
	\label{fig:SIWIMP}
\end{figure}

Figure \ref{fig:SIWIMP} shows the spin-independent WIMP-nucleus scattering cross section vs the 2016 LUX experimental limits \cite{Akerib:2016lao,Akerib:2016vxi}. One can see that the vast majority of parameter space is not yet touched by DD, and nearly all of the red sneutrino points are far from the near-future experimental reach.

\begin{figure}[h]
	\centering
	\includegraphics[width=0.9\linewidth]{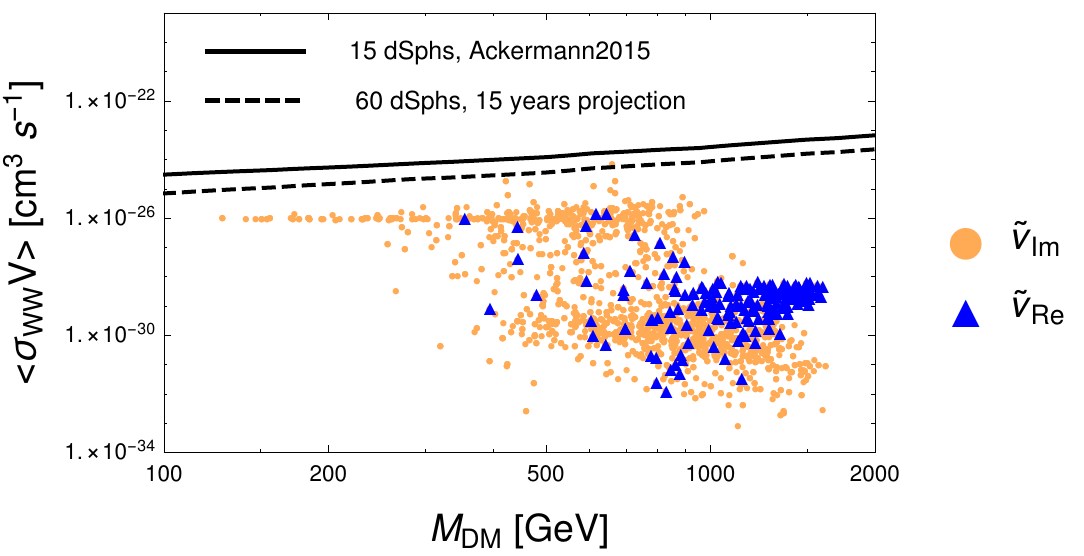}
	\caption{Thermal cross section for DM DM $\to W^+W^-$ annihilation as predicted by theory as a function of the DM mass, for  
		CP-even (blue) and CP-odd (orange) sneutrinos. Also shown are the FermiLAT limit from dSphs at present (solid black) and as projection for 15 years from now (dashed black).
		All points obey the relic density upper limit, for which rescaling, where necessary, has been applied.}
	\label{fig:sigmav_ww_15y60dSphsProj_relic}
\end{figure}

\begin{figure}
	\centering
	\includegraphics[width=0.8\linewidth]{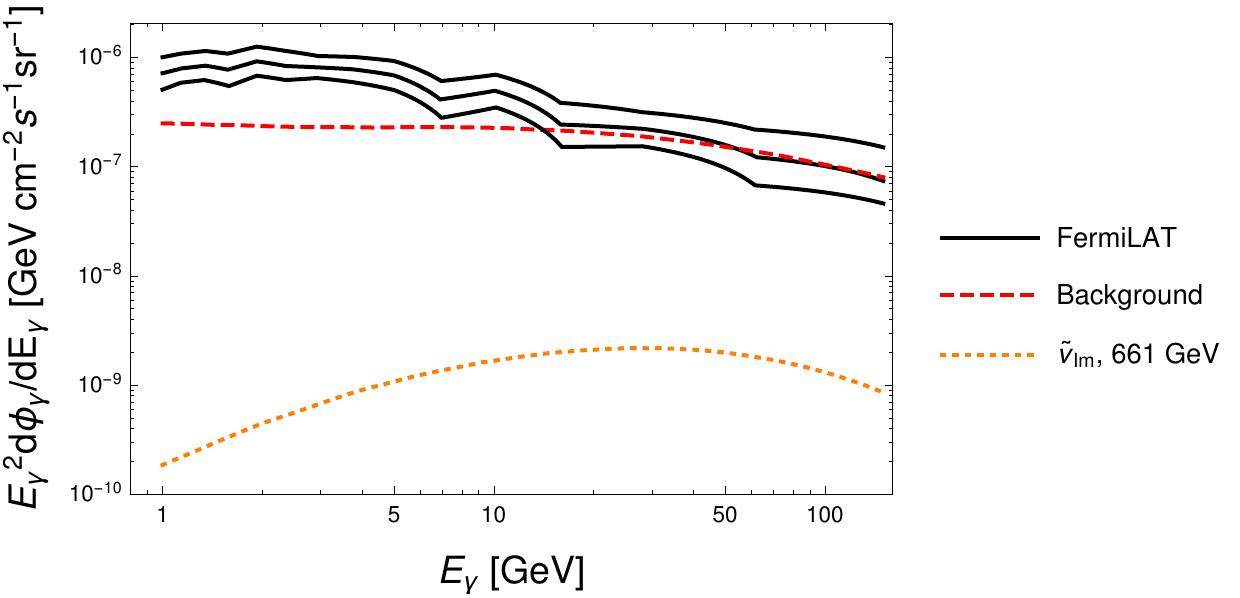}
	\caption{Differential flux of $\gamma$-ray secondary radiation induced by DM DM $\to W^+W^-$ annihilation as a function of the photon energy, with fixed DM mass,
		for our benchmark  CP-odd sneutrino (orange). The corresponding distribution for the background is also given  (red). The FermiLAT present data (with error) are in black.
		The sneutrino point considered is compliant with the relic density constraint taken as an upper limit.}
	\label{fig:FermilatE2DNDE}
\end{figure}

\begin{figure}
	\centering
	\includegraphics[width=0.9\linewidth]{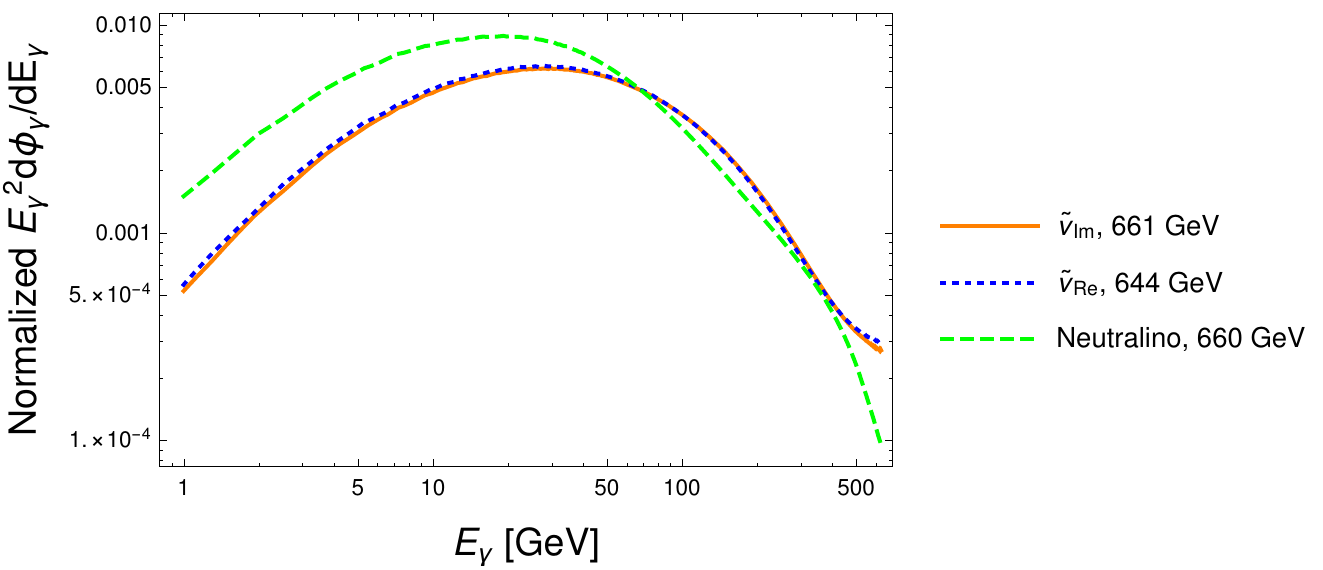}
	\caption{Differential flux of $\gamma$-ray secondary radiation induced by DM DM $\to W^+W^-$ scatterings as a function of the photon energy, with fixed DM mass,
		for  our benchmark CP-even (blue) and CP-odd (orange) sneutrinos. The corresponding distribution for a neutralino is also given for comparison (green). Normalisation is the same for all curves.}
	\label{fig:NormalizedComparison650E2LogLog}
\end{figure}

Another method of detecting sneutrino DM comes from indirect detection. Here annihilations to charged $W^+W^-$ pairs can then radiate photons in the galactic centre, which may be observed by the Fermi-LAT experiment. In figure \ref{fig:sigmav_ww_15y60dSphsProj_relic} we plot the self-annihilation cross section in the charged $W^+W^-$ channel vs the DM mass for both CP-even and CP-odd sneutrinos. We overlay the current Fermi-LAT limit, which uses 15 dwarf Spheroidal galaxies (dSphs) \cite{Ackermann:2015zua}. The projected sensitivity over the next 15 years is also drawn, which extends the search via observation of 60 dSphs samples. A single GUT-constrained parameter point exists which is detectable according to the projected limits. This signals that there are regions of parameter space which are beginning to be excluded over the next few years of observation. In addition to the total integrated flux measurement, one may also study the differential $\gamma$-ray flux due to sneutrino annihilations in the centre of the Milky Way. This is done in figure \ref{fig:FermilatE2DNDE}, where we show the differential $\gamma$-ray flux vs the energy of photons for a particular CP-odd sneutrino LSP DM candidate. We also plot the background and Fermi-LAT data, with errors. The signal from the sneutrino is well below the background, so one may not see a large signal at particular photon energy. We notice here, though, that the limiting photon energy is by an experimental upper limit of $\sim 300$ GeV, and the sneutrino is capable of emitting $\gamma$-rays of energies up to its mass, $661$ GeV. In the future, when one is able to detect higher photon energies, there may be a characteristic signal emerge. Should such a signal be observed, one may wonder whether this could disentangle the CP-odd and CP-even case with other DM candidates. We draw the spectrum shape for these three candidates in figure \ref{fig:NormalizedComparison650E2LogLog}. One can see that the CP-odd and CP-even shapes look identical, so one may not disentangle these two scenarios. However, this does look characteristically different from the neutralino shape. We speculate that this could be to do with the spin of the candidate, where the sneutrino is a scalar and the neutralino is a fermion.

\section{LHC Searches}
\label{sec:Collider_Searches}
In addition to direct and indirect detection, one can also search for sneutrino DM via LHC searches. There is little mixing between the left and RH sneutrinos, as the Yukawa coupling is set to be very small $(Y_\nu \sim \mathcal{O}(10^{-6}))$ to obtain the correct LH neutrino masses with an $\mathcal{O}(1)$ TeV RH neutrino scale. So there is no hope to produce the DM via $W^\pm$ or $Z$ with any appreciable cross section. Given this, we have performed a general search using the MadGraph program \cite{Alwall:2014hca} using several benchmark sneutrino points, both of CP-even and CP-odd varieties. The most well-established DM search is via mono-jet, whereby pair production of DM is detected by initial radiation of a single gluon, which will show the signature of a mono-jet plus missing energy. However, this does not uniquely identify the BLSSM whatsoever. A more unique direct DM production signature could be through $Z'$ decay, in the process  $pp \rightarrow Z' \rightarrow \tilde{\nu}_{LSP} \tilde{\nu}_{{NLSP}}$, where the NLSP is forced to decay to the LSP via a SM Z decay, despite the small coupling. However, as the mass of our $Z'$ is fixed to be 4 TeV, so as to comply with dilepton searches, the cross section for this process is too small to observe, $\sigma \simeq 0.025$ for both CP-even and CP-odd LSP benchmark points. Another possibility to search for sneutrino DM at the LHC can be through slepton pair production, with cross sections $\sim$ 0.1 fbs. For a light slepton mass, the only possible decay path can be through $\tilde l \rightarrow W^\pm \tilde{\nu}_{LSP}$, despite the suppressed Yukawa coupling, yielding a dilepton signature. For a different mass configuration, one may find the more complicated decay chain $\tilde l \rightarrow \tilde \chi^0 l$ with $\tilde \chi^0 \rightarrow \nu_h \tilde{\nu}_{LSP}$, where $\nu_h$ is the heavy neutrino. The latter will mainly undergo $\nu_h \rightarrow W^\pm l^\mp$ or $\nu_h \rightarrow Z \nu_l$ decay, thus providing fully or semi-leptonic signatures which would be very distinctive. Finally, one may also consider signatures from squark pair production, which can offer some of the largest cross sections $(\mathcal{O}($fbs). The typical decay chain is now through $\tilde t \rightarrow \tilde \chi^0 \, t$, with large branching fractions. The neutralino may decay as before, $\tilde \chi^0 \rightarrow \nu_h \tilde{\nu}_{LSP}$. This signature would involve several jet plus multi-lepton final states.

\section{Conclusion}
\label{sec:Conclusion}
We have compared naturalness and DM properties of the BLSSM and MSSM. We have seen that both models possess similar levels of FT but, when one includes the requirement of an LSP which satisfies the DM bounds, the BLSSM has a far larger region of available parameter space, due to the unique RH sneutrino. We then went on to assess the ability to detect this DM candidate, and saw that whilst direct detection in the near future seems unlikely, both CP-even and CP-odd candidates may be observed via indirect detection in the next few years. Finally, we identified several interesting LHC signatures which, if seen, would provide strong evidence for sneutrino DM.

\acknowledgments
SM is supported in part through the NExT Institute. The work of LDR has been supported by the STFC/COFUND Rutherford International Fellowship scheme. The work of CM is supported by the `Angelo Della Riccia' foundation and by the Centre of Excellence project No TK133 `Dark Side of the Universe'. The work of SK is partially supported by the STDF project 13858. SK, SJDK and SM acknowledge support from the grant H2020-MSCA-RISE-2014 n. 645722 (NonMinimalHiggs). SJDK and SM acknowledge support from the STFC Consolidated grant ST/L000296/1.  SuK is supported by the `New Frontiers' program of the Austrian Academy of Sciences and by FWF project number V592-N27.

\end{document}